\documentclass[prb,nofootinbib,twocolumn,superscriptaddress]{revtex4} 

\usepackage{graphicx}
\usepackage{dcolumn}
\usepackage{bm}
\usepackage{threeparttable}
\usepackage{times}
\usepackage{mathptmx}
\usepackage{lscape}
\usepackage{natbib}
\usepackage{amsmath}
\usepackage{amssymb}
\usepackage{braket}
\usepackage{comment}
\usepackage{color}

\def\degree{\kern-.2em\r{}\kern-.3em}

\begin{document}

\title{ Geometric Study on Noncommutativity in Canonical Nonlinearity  }

\author{Atsuro Arai}
\affiliation{
Department of Materials Science and Engineering,  Kyoto University, Sakyo, Kyoto 606-8501, Japan\\
}%

\author{Seiru Miyagawa}
\affiliation{
Department of Materials Science and Engineering,  Kyoto University, Sakyo, Kyoto 606-8501, Japan\\
}%

\author{Koretaka Yuge}
\affiliation{
Department of Materials Science and Engineering,  Kyoto University, Sakyo, Kyoto 606-8501, Japan\\
}%

\begin{abstract}
{ 

For classical discrete systems under constant composition, canonical average provides equilibrium configuration from a set of many-body interactions, which typically acts as nonlinear map (so-called ``canonical nonlinearity''). The nonlinearity has recently been investigated in terms of configurational geometry (i.e., geometric information about configurational density of states: CDOS), where two measures for the nonlinearity as vector field on configuration space and divergence on statistical manifold are introduced on individual configuration. 
Then, concepts of these measures are further unified through stochastic thermodynamic treatment, which enables formulation of the nonlinearity on multiple configrations. 
While these studies provide deeper insight into understanding the nonlinearity, thermodynamic treatment for non-separability in structural degrees of freedom (NS), which is a partial and non-trivial contribution to the nonlinearity, still remains difficult due mainly to nonexistence of the corresponding CDOS. Our recent study partially overcome the problem, by considering additional information of ``non-commutativity'' for the nonlinearity (NC), where the sum of non-separability and non-commutativity can be fully handled by the thermodynamic treatment. 
The present study focuses on clarfying the basic behavior of the NC, in terms of the geometric information about configurational geometry that is much more easily-accesible information than NC. For this purpose, we here employ the coarse-grained state model (CSM) that can qualitatively capture the nonlinearity information through the coarse-grained configuration space with tractable numbers of parameters, where the present CSM can implicitly include the difference in coordination numbers on real lattice. 
We find that (i) the NC exhibit positive correlation with asymmetric Hausdorff distance in configurational polyhedra between of pracitcal and separable systems in terms of SDFs, especially for partically-ordered-state-rich region, (ii) the NC take lower value when the difference in coordination number becomes small, indicating that we can obtain better thermodynamic bound for NS in such systems, and (iii) the positive correlation can be further modified when we additionally include information about non-separability in the practical CDOS. 

}
\end{abstract}


\maketitle

\section{Introduction}

When we consider classical discrete systems with constant composition described by $f$ SDFs on given lattice, configuration along the $p$-th coordination (among $\left\{ q_{1},\cdots, q_{f} \right\}$) in thermodynamic equilibrium can be expressed through the well-known canonical average, namely, 
\begin{eqnarray}
\label{eq:can}
\Braket{ q_{p}}_{Z} = Z^{-1} \sum_{i} q_{p}^{\left( i \right)} \exp \left( -\beta U^{\left( i \right)} \right),
\end{eqnarray}
where $Z$ denotes partition function, $\beta$ inverse temperature, $U$ potential energy, and summation is taken over all possible configurations $i$. 
When we employ the complete basis functions for e.g., the generalized Ising model,\cite{ce} $U$ for configuration $k$ can be rigorously written by $U^{\left( k \right)} = \sum_{j=1}^{f}\Braket{U|q_{j}}q_{j}^{\left( k \right)}$, where $\Braket{\quad|\quad}$ denotes inner product, i.e., trace over possible configurations 
Based on the description, canonical average of Eq.~\eqref{eq:can} can now be interpreted as a map $\phi$ of $\phi: \vec{U}\mapsto \vec{Q}_{Z}$, where $\vec{U}=\left( \Braket{U|q_{1}},\cdots, \Braket{U|q_{f}} \right)$ and $\vec{Q}_{Z}=\left( \Braket{q_{1}}_{Z},\cdots, \Braket{q_{f}}_{Z} \right)$, which generally exhibits a complicated nonlinearity for practical systems (such as substitutional alloys).

Especially for alloy thermodynamics, the nonlinearity has been \textit{implicitly} investigated through a plethora of theoretical approaches to predict their equilibrium properties, based on e.g., the Metropolis algorithm, multihistgram method, multicanonical ensemble, and entropic sampling for effective exploration of the target space.\cite{mc1,mc2,mc3,mc4} They are further combined with the generalized Ising model augmented with a variety of optimization technique for many-body interactions, including cross-validation, genetic algorism and regression in machine learning.\cite{cm1,cm2,cm3,cm4,cm5,cm6} 
However, they do not essentially provide the nature of the nonlinearity, especially from the perspective of ``configurational geometry'' based on the information about configuraional density of states (CDOS) independent of many-body interaction as well as temperature.

In order to address the issues, we have recently introduce two measures for the nonlinearity in terms of the configurational geometry: The one is the vector field on configuration space,\cite{asdf} which provides local information about the nonlinearity at a given configuration, and the another is the Kullback-Leibler (KL) divergence on statistical manifold, which can include further non-local information around the given configuration.\cite{ig} We then further develop thermodynamic treatment of the nonlinearity by transforming the system transition driven by nonlinearity into the transition by heat transfer and external work: The treatment enables unifying the concept of two different measures for the nonlinearity, and formulating the averaged nonlinear character over multiple configurations.\cite{nol-thermo}

Although these studies provide deeper insight into the nonlinearity based on configurational geometry, thermodynamic treatement for partial, non-trivial contribution to the nonlinearity, the so-called ``nonseparability in structural degrees of freedom'' (NS) still remains essentially difficult due mainly to the non-existence of the unique CDOS for NS. Very recently, the problem has been effectively addressed by introducing additional information about the nonlinearity, i.e., ``non-commutativity'' (NC) between canonical map and separable-operation for SDFs: The sum of NS and NC can be reasonablly handled through the previously-proposed thermodynamic approach.\cite{nc} The remaining problem is that the NC is difficult to straightforwardly address due mainly to requiring direct information about CDOS. 
Therefore, at the starting point, phenomenological study on the NC for variety of landscapes of CDOS, and introducing easily-accesible measure for effectively capturing the characteristics of the NC on multiple CDOS, is highly desired.

The present study address these points, by analyzing the NC for the coarse-graind state model (CSM) for configuration space to construct CDOS with a few parameters, and proposing (assymetric) Hausdorff distance in configurational polyhedra (CP) as an measure for the NC. We find that (i) difference in pair coordination number and average NC over configuratoins exhibit strong positive, almost-linear correlation, and (ii) Hasdorff distance and the NC exhibit reasonable possitive correlation for restricted set of CDOS with typically-holded condition for practical systems. These results certainly indicates that better bound for averaged NS in thermodynamic treatment is achieved for smaller difference in coordination number and for the system with geometry of CP is close to the hyperrectangular obtained through direct products of the CP for individual SDF.

\section{Concept and Discussion}
\subsubsection*{Nonlinearity Measure and Decomposition }
We first briefly explain the two measures for nonlinearity on individual configuration $\vec{q}$. The first one is the vector field $\vec{H}$ on configuration space, defined by 
\begin{eqnarray}
\label{eq:asdf}
\vec{H}\left( \vec{q} \right) = \left\{ \phi\left( \beta \right)\circ \left( -\beta\cdot \Gamma \right)^{-1} \right\}\cdot \vec{q} - \vec{q}.
\end{eqnarray}
Here, $\circ$ represent composite map and $\Gamma$ covariance matrix for CDOS. The main characteristcs for the vector field are (i) $\vec{H}$ is independent of many-body interaction and temperature, and (ii) $\vec{H}$ takes zero-vector when $\phi$ is locally linear around $\vec{q}$, which therefore capture the local nonlinear information at given configuration. Then hhe concept of $\vec{H}$ can be straightforwardly extended to another measure based on KL divergence, i.e., corresponding nonlinearity on configuration $\vec{q}$ is defined by\cite{ig} 
\begin{eqnarray}
D_{\textrm{NOL}}^{q} = D\left( P^{\textrm{E}} : P^{\textrm{G}} \right), 
\end{eqnarray}
where $P^{\textrm{E}}$ and $P^{\textrm{G}}$	repsectively denotes canonical distribution obtained from the following interaction:
\begin{eqnarray}
V = \left( -\beta\cdot\Gamma \right)^{-1}\cdot \vec{q}.
\end{eqnarray}
with pracitcal and Gaussian CDOS (with the same covariance matrix of $\Gamma$). 
We note that  $D_{\textrm{NOL}}^{q}$ is also independent of the temperature and many-body interactions, which is a common characteristic with $\vec{H}$. 

Based on the information geometry, for binary system with pair correlations, we clarify that there exist partial, non-trivial contribution to the nonlinearity $D_{\textrm{NOL}}$: This corresponds to the ``nonseparabirity'' (NS) discribed above, given by
\begin{eqnarray}
D_{\textrm{NS}} = D_{\textrm{KL}}\left( P^{\textrm{S}}:P^{\textrm{G}} \right).
\end{eqnarray}
Here, $P^{\textrm{S}}$ denotes taking separable-operation $\Psi$ on $P^{\textrm{E}}$ (i.e., taking its marginal distribution and then their product).
In the above equation, NS purely capture the non-local contribution which is not included in $\vec{H}$: Therefore, NS can be non-trivial contribution to the nonlinearity, where the present study (indirectly) focuses on. 

\subsubsection*{Noncommutativity for Nonlinearity}
From the definition of NS, it is now clear that the corresponding CDOS cannot be uniquely determined (or not exist), which makes essentially difficult in its thermodynamic treatment. To address the problem, we recently introduce additional contribution called ``noncommutativity'' in the nonlinearity $D_{\textrm{NC}}$, defined by
\begin{eqnarray}
D_{\textrm{NC}} = D_{\textrm{KL}}\left( P^{\textrm{S}} : P^{\textrm{S2}} \right),
\end{eqnarray}
where $P^{\textrm{S2}}$ represents canonical distribution obtained from the CDOS of $\Psi\cdot g $. 
The definition of NC therefore measure the noncommutative character between $\Psi$ and $\phi$ on given probability distribution $P$ of
\begin{eqnarray}
\label{eq:nonc}
\left[ \Psi, \phi \right] P = \Psi\cdot\phi\cdot P - \phi\cdot\Psi\cdot P \neq 0
\end{eqnarray}
as KL divergence.
Then it is clear that (i) $P^{\textrm{S}}$ and $P^{\textrm{S2}}$ sits on the common $e$-flat subspace in statistical manifold, and (ii) $P^{\textrm{S2}}$ can have unique CDOS of $\Psi\cdot g$, whose thermodynamic treatment is straightforward. From the fact of (i), we can see the relationship from the genelarized Pythagorean theorem:
\begin{eqnarray}
D_{\textrm{NS}} = D_{\textrm{NS2}} - D_{\textrm{NC}},
\end{eqnarray}
where $D_{\textrm{NS2}}$ corresponds to \textit{pseudo} NS for artificially prepared CDOS of $\Psi\cdot g$. 

Since again, thermodynamic treatment for $D_{\textrm{NS2}}$ is easily handled, analyzing the behavior of $D_{\textrm{NC}}$ becomes significant for (indirect) thermodynamic treatment for $D_{\textrm{NS}}$. 

\subsubsection*{Coarse-Grainded State Model}
From the definition of NC, it appears difficult to construct a set of configuration to mimic the corresponding CDOS of $\Psi\cdot g$, which implicitly means that NC requires direct information about the original CDOS landscape even for merely taking its canonical distribution of $P^{\textrm{S2}}$, which is typically intractable for practical substituional alloys. 

Therefore, in order to effectively capture the basic behavior of NC in terms of CDOS landscape, we here introduce the coarse-grained state model (CSM) for coordinates of $\left( q_{1}, q_{2} \right)$, i.e., $f=2$ SDFs system, where the configuration space is represented by coarse-grained $3\times 3 = 9$ states: Each SDF exhibit 3 discrete state with correlation function respectively taking $\left\{ -1, 0, 1 \right\}$. Note that the idea in CSM of employing $3\times 3$ states is validated through (i) distinct feature of ASDF in 9 areas in configurationspace for practical fcc and bcc binary alloys with variety of sets of pair correlations, and (ii) dynamic mode decomposition of changes in ASDF w.r.t. changes in coordination numbers for artificially constructed coupled-linear systems. For practical binary systems with pair correlations, CDOS landscape typically strongly depends on its (difference in )constituent coordination numbers, and covariance matrix $\Gamma$ should be diagonal at thermodynamic limit: When we employ these conditions as well as center of gravity of the CDOS taking origin, we can introduce corresponding CDOS for CSM as follows:
\begin{eqnarray}
\label{eq:csm-cdos}
g=\left( 
\begin{array}{ccc}
x & y+z & x \\
y-z & 1-4x-4y & y-z \\
x & y+z & x
\end{array} 
\right).
\end{eqnarray}

\begin{figure}[h]
\begin{center}
\includegraphics[width=0.6\linewidth]{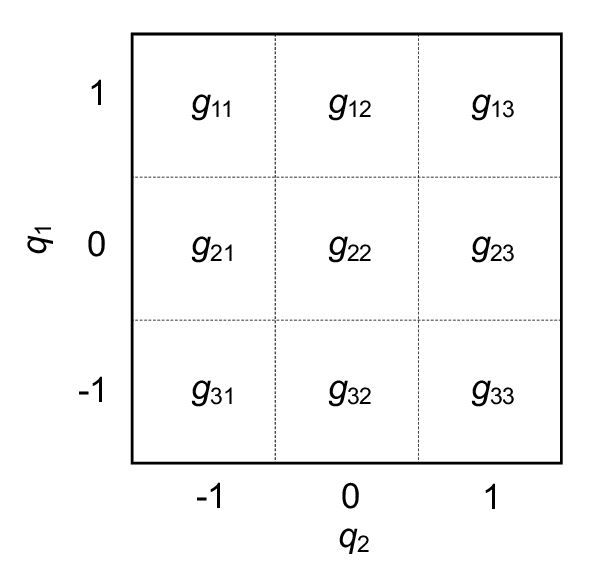}
\caption{ Schematic illustration of coarse-graind configuration space of $f=2$ SDFs, with corresponding CDOS.}
\label{fig:conf-space}
\end{center}
\end{figure}

With this model, elements of $g_{11}$, $g_{13}$, $g_{31}$, $g_{33}$ corresponds to around ordered states (i.e., vertices of configurational polyhedra), $g_{12}$, $g_{21}$, $g_{23}$, $g_{32}$ to partially ordered states, and $g_{22}$ to random state, which is schematically illustrated in Fig.~\ref{fig:conf-space}. 

Variance of marginal CDOS along $q_{1}$ and $q_{2}$ is respectively given by $ 4x + 2y + 2z $ and $4x + 2y -2z$. It is known that under constant system size and composition, variance of practical CDOS is proportional to inverse of pair coordination number, indicating that parameter $z$ in CSM has role of controlling the difference in coordination number.  $x$ and $y$ individually controll the  number of ordered (i.e. ground) states and partially ordered states, clearly seen from Eq.~\eqref{eq:csm-cdos}.  When we further consider the practical condition that CDOS around random states is higher than other states, we set $0\le x \le 0.2$ and $0\le y \le 0.2$. 

\subsubsection*{Introducing Easily-Accesible Measure}
As discussed, since the NC directly requires information about landscape of CDOS, whose estimation is typically difficlut for pracitcal systems, we here introduce alternative approach of CSM. In this sense, therefore, another appropriate measure for the NC to connect configrational geometry of CSM and that of practical system, which is easily accesible for practical system, would be highly desired. 

From Eq.~\eqref{eq:nonc}, it is expected that the NC strongly reflects the changes in geometry of configurational polyhedra from that for separable SDFs system, since range for $\phi\cdot\Psi\cdot P$ takes within the practical CP while that for $\Psi\cdot\phi\cdot P$ can take out of the range. 

With these considerations, we here introduce \textit{effective}, asymmetric Hausdorff distance between configurational polyhedra of SDFs-separable (CP0) and that of CSM (CP), defined by 
\begin{eqnarray}
R_{\textrm{H}} = \underset{a\in \textrm{CP0}}{\textrm{sup}}\left\{ \underset{b\in \textrm{CP}}{\textrm{inf}} d\left( a,b \right) \right\},
\end{eqnarray}
where $d$ is Euclidean distance. The reason for employing ``asymmetric'' $R_{\textrm{H}}$ is that CP0 takes constant geometry while CP does not. 

For the present CSM, we assume that (i) $R_{\textrm{H}}$ depends only on the CDOS around ground-state, and (ii) $R_{\textrm{H}}$ is the function of $x^{0.5}$ from the dimension analysis. When we take boundary condition of $R_{\textrm{H}}=0$ at $x=0.2$ and $R_{\textrm{H}}=\sqrt{2}/2$ at $x=0$, we provide
\begin{eqnarray}
R_{\textrm{H}} = -\frac{\sqrt{10}}{2} x^{0.5} + \frac{\sqrt{2}}{2}.
\end{eqnarray}
Note that this boundary condition corresponds to the situation that CP takes the same geometry as CP0 at maximum $x$ of 0.2, and CP at ordered states take isosceles right triangles at minimum $x$ of zero. 

\subsubsection*{Landscape of Noncommutativity}
\begin{figure}[h]
\begin{center}
\includegraphics[width=0.98\linewidth]{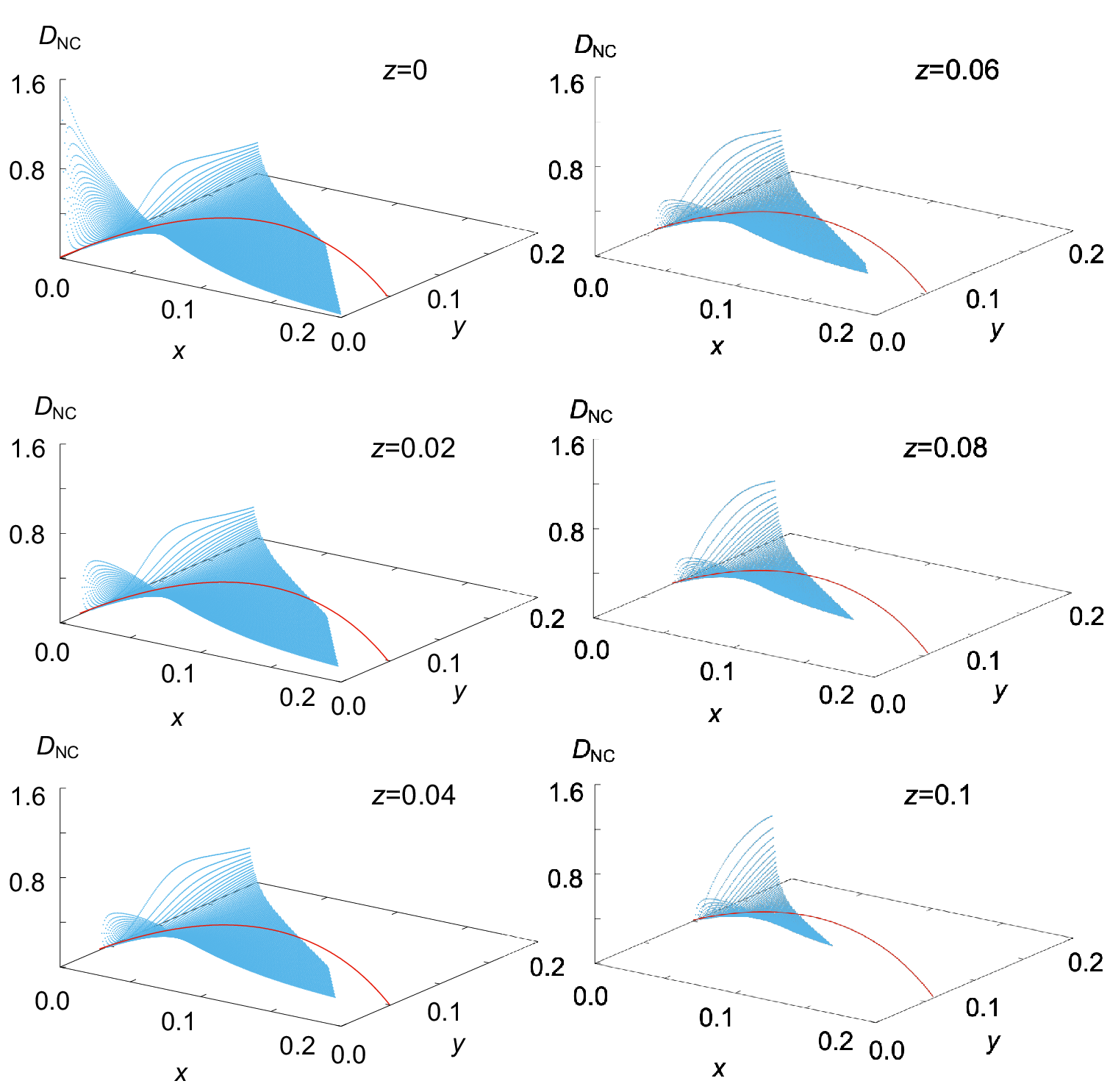}
\caption{ Landscape of averaged noncommutativity $D_{\textrm{NC}}$ over possible 9 states as a functions of $x$ and $y$ at individual $z$ value. Curve at bottom surface represents condition for $x$ and $y$ where CDOS becomes separable. }
\label{fig:dnc-xy}
\end{center}
\end{figure}

We first show the CDOS-landscape dependence of averaged noncommutativity over the nine states at individual $z$, given in Fig.~\ref{fig:dnc-xy}.
From the definition of noncommutativity, we can reasonably see that when CDOS becomes separable (curves at bottom surface), $D_{\textrm{NC}}$ takes zero. Landscape of the CDOS exhibits significant asymmetry across the separable curve: In front of the curve (hereinafter called $x$-rich region), $D_{\textrm{NC}}$ takes maximum at lower $x$ and $y$, while behind the curve (hereinafter called $y$-rich region), $D_{\textrm{NC}}$ takes maximum at lower $x$ and higher $y$. 

\begin{figure}[h]
\begin{center}
\includegraphics[width=0.99\linewidth]{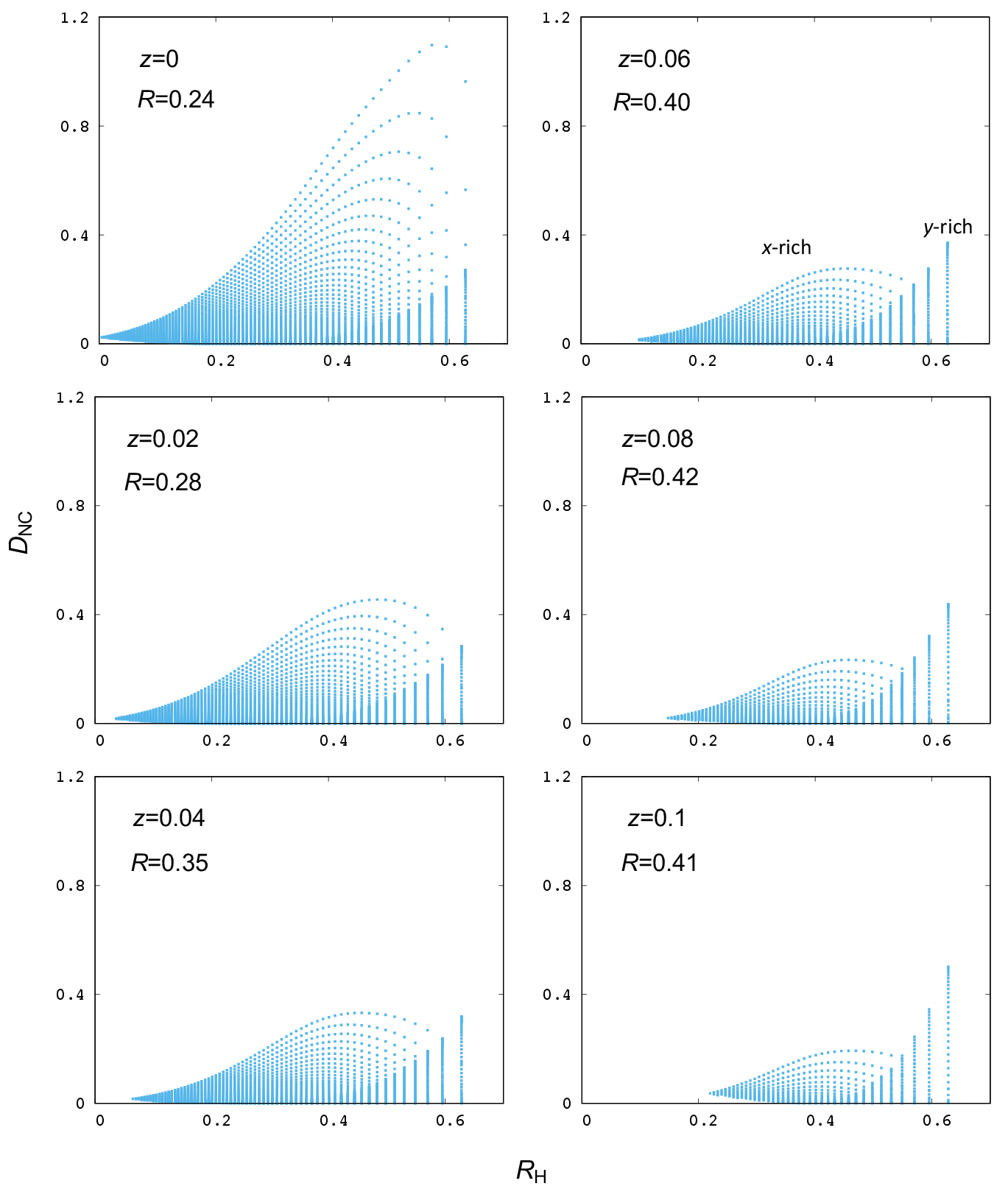}
\caption{ Correlation between averaged noncommutativity $D_{\textrm{NC}}$ and the Hausdorff distance $R_{\textrm{H}}$ at individual $z$ value, where correlation coefficient $R$ is given together.}
\label{fig:dnc-rh}
\end{center}
\end{figure}

Next, we confirm the correlation between the averaged $D_{\textrm{NC}}$ and the introduced Hausdorff distance $R_{\textrm{H}}$, shown in Fig.~\ref{fig:dnc-rh}.
From the figure, we can clearly see (i) a slightly weak, positive correlation: increasing $R_{\textrm{H}}$ typically tends to increase 
the averaged noncommutativity $D_{\textrm{NC}}$, and (ii) there exhibits two distinct correlations at $x$-rich and $y$-rich region (given for example at right, topmost figure in Fig.~\ref{fig:dnc-rh}), which can be reasonablly seen from the $D_{\textrm{NC}}$ landscape of Fig.~\ref{fig:dnc-xy}. 
However, the minimum $D_{\textrm{NC}}$ still exhibit around zero for every $R_{\textrm{H}}$, certainly indicating that information about $R_{\textrm{H}}$ is not even sufficient to qualitatively capture the $D_{\textrm{NC}}$ landscape. 

\subsubsection*{Modified Measure for Noncommutativity}
From above discussions, further information would be highly required to interpret the characteristic landscape of $D_{\textrm{NC}}$. From Fig.~\ref{fig:dnc-xy}, we can reasonably expect that landscape of $D_{\textrm{NC}}$ is strongly bounded by the existence of separability condition, where $D_{\textrm{NC}}$ should exactly take zero at the condition. 

Since this condition is originally derived from non-separability in CDOS of the CSM, we here employ the combination of information about $R_{\textrm{H}}$ and $D_{\textrm{NS}}$ in the CDOS, as the following product form:
\begin{eqnarray}
R_{\textrm{H}}\cdot\sqrt{D_{\textrm{NS}}}
\end{eqnarray}
for a new measure of the $D_{\textrm{NC}}$ landscape. The reason why taking square root of $D_{\textrm{NS}}$ is that KL divergence corresponds to (pseudo) squared distance on statistical manifold, where the introduced measure composed of $R_{\textrm{H}}$ as Euclidean distance and $\sqrt{D}_{\textrm{NS}}$ as (pseudo) distance on statistical manifold: Note that when we replace $\sqrt{D_{\textrm{NS}}}$ into $D_{\textrm{NS}}$, the following discussion for the correlation between $D_{\textrm{NC}}$ and the measure does not essentially change. 

\begin{figure}[h]
\begin{center}
\includegraphics[width=0.7\linewidth]{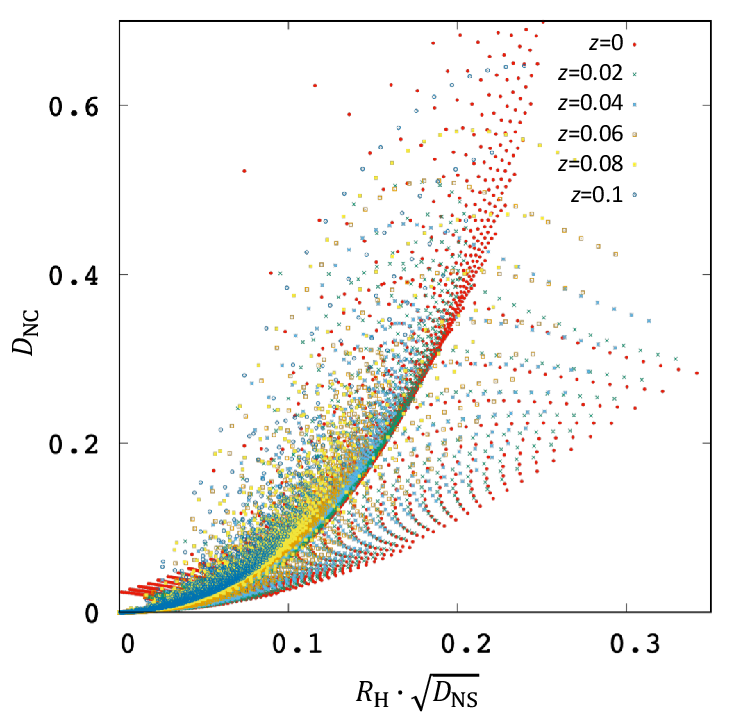}
\caption{ Correlation between noncommutativity and newly introduced measure of $R_{\textrm{H}}\cdot \sqrt{D_{\textrm{NS}}}$. }
\label{fig:dnc-rh-dns}
\end{center}
\end{figure}

We show in Fig.~\ref{fig:dnc-rh-dns} the resultant correlation between noncommutativity and the introduced measure.
We can see that the new measure exhibits more reasonable, positive correlation with $D_{\textrm{NC}}$ where its minimum at each $z$ monotonically increases with increasing the measure, which also holds for their averaged value of $D_{\textrm{NC}}$ over possible $x$ and $y$ values. This strongly indicates that to capture the characteristics of NC, geometric information about area-constraint for correlation functions as well as non-separable character for practical CDOS should be essentially required.

\section{Conclusions}
Based on employing coarse-graind configuration space model (CSM), we investigate basic behavior of noncommutativity (NC), which is required for thermodynamic treatment of the non-trivial, non-local contributin to the nonlinearity in canonical map. We find that the NC can be reasonablly characterized by joint information about geometric distance in configurational polyhedra between practical and separable system in terms of SDFs, and about nonseparability in the practical CDOS, which exhibit clear possitive correlation.

\section{Acknowledgement}
This work was supported by Grant-in-Aids for Scientific Research on Innovative Areas on High Entropy Alloys through the grant number JP18H05453 and  from the MEXT of Japan, and Research Grant from Hitachi Metals$\cdot$Materials Science Foundation.


\begin{thebibliography}{9}
\bibitem{ce} J.M. Sanchez, F. Ducastelle, and D. Gratias, Physica A \textbf{128}, 334 (1984).
\bibitem{mc1} N. Metropolis, A. W. Rosenbluth, M. N. Rosenbluth, A. H. Tellerand, and E. Teller, J. Chem. Phys. \textbf{21}, 1087 (1953).
\bibitem{mc2} A.M. Ferrenberg and R. H. Swendsen, Phys. Rev. Lett. \textbf{63}, 1195 (1989).
\bibitem{mc3} G. Bhanot, R. Salvador, S. Black, P. Carter, and R. Toral, Phys. Rev. Lett. \textbf{59}, 803 (1987).
\bibitem{mc4} J. Lee, Phys. Rev. Lett. 71, 211 (1993).
\bibitem{cm1} V. Blum, G. L. W. Hart, M. J. Walorski, and A. Zunger, Phys. Rev. B \textbf{72}, 165113 (2005).
\bibitem{cm2} A. Seko, Y. Koyama, and I. Tanaka, Phys. Rev. B \textbf{80}, 165122 (2009).
\bibitem{cm3} T. Mueller and G. Ceder, Phys. Rev. B \textbf{82}, 184107 (2010).
\bibitem{cm4} K. Yuge, Phys. Rev. B \textbf{85}, 144105 (2012).
\bibitem{cm5} L.J. Nelson, G.L.W. Hart, F. Zhou,and V. Ozolins, Phys. Rev. B \textbf{87}, 035125 (2013).
\bibitem{cm6} A.R. Natarajan and A. Van der Ven, NPJ Comput. Mater. \textbf{4}, 56 (2018).
\bibitem{asdf} K. Yuge, J. Phys. Soc. Jpn. \textbf{86}, 104802 (2018).
\bibitem{ig} K. Yuge, J. Phys. Soc. Jpn. \textbf{91}, 014802 (2022).
\bibitem{nol-thermo} K. Yuge, J. Phys. Soc. Jpn. \textbf{93}, 094802 (2024).
\bibitem{nc} K. Yuge, arXiv:2303.16311 [cond-mat.stat-mech].
Mathematical Physics \textbf{16}, Springer, 2022). 
\end{thebibliography}
\end{document}